\documentclass[]{aiaa-tc}

\usepackage[T1]{fontenc}
\usepackage[english]{babel}
\usepackage{textcomp}

\usepackage{amsmath, amssymb}
\usepackage{graphicx, color, xcolor, wrapfig, subcaption}
\usepackage{listings}

\usepackage[hyphens]{url}
\usepackage[breaklinks=true, linkcolor=blue, citecolor=blue, colorlinks=true]{hyperref}
\usepackage{booktabs}

\usepackage[retainorgcmds]{IEEEtrantools}

\usepackage{nicefrac}

\usepackage[binary-units]{siunitx}

\definecolor{dkgreen}{rgb}{0,0.6,0}
\definecolor{gray}{rgb}{0.5,0.5,0.5}
\definecolor{mauve}{rgb}{0.58,0,0.82}

\usepackage{nomencl}
\makenomenclature

\def\CC{{C\nolinebreak[4]\hspace{-.05em}\raisebox{.4ex}{\footnotesize ++}}}

\usepackage{todonotes}
\graphicspath{ {images/} }

\title{An initial investigation of the performance of GPU-based swept time-space decomposition}

\author{
	Daniel Magee%
	\thanks{Graduate Research Assistant, School of Mechanical, Industrial, and Manufacturing Engineering; AIAA Student Member}
	\ and Kyle E.~Niemeyer\thanks{Assistant Professor, School of Mechanical, Industrial, and Manufacturing Engineering; AIAA Professional Member}\\
	{\normalsize\itshape{Oregon State University, Corvallis OR 97331-6001, USA}}
}

\AIAApapernumber{YEAR-NUMBER}
\AIAAconference{Conference Name, Date, and Location}
\AIAAcopyright{\AIAAcopyrightD{YEAR}}

\begin{document}

\maketitle

\begin{abstract}

Simulations of physical phenomena are essential to the expedient design of precision components in aerospace and other high-tech industries.
These phenomena are often described by mathematical models involving partial differential equations (PDEs) without exact solutions.
Modern design problems require simulations with a level of resolution that is difficult to achieve in a reasonable amount of time even in effectively parallelized solvers.
Though the scale of the problem relative to available computing power is the greatest impediment to accelerating these applications, significant performance gains can be achieved through careful attention to the details of memory accesses.
Parallelized PDE solvers are subject to a trade-off in memory management: store the solution for each timestep in abundant, global memory with high access costs or in a limited, private memory with low access costs that must be passed between nodes.
The GPU implementation of swept time-space decomposition presented here mitigates this dilemma by using private (shared) memory, avoiding internode communication, and overwriting unnecessary values.
It shows significant improvement in the execution time of the PDE solvers in one dimension achieving speedups of \num{6}-\num{2}$\times$ for large and small problem sizes respectively compared to naive GPU versions and \num{7}-\num{300}$\times$ compared to parallel CPU versions.

\end{abstract}

\section{Introduction}

High-fidelity computational fluid dynamics (CFD) simulations are essential in the development of aerospace technologies such as rocket launch vehicles and jet engines.
The motivation for this project is to accelerate such simulations to approach real-time execution, simulation at the speed of nature, in accordance with the high-performance computing development goals set out in the CFD Vision 2030 report~\cite{slotnick:2014}.

Classic approaches to discrete decomposition of explicit, time-stepping partial differential equations (PDEs) incur high computational performance costs by communicating between nodes every timestep in a parallel distributed memory system.
This communication cost is composed of two parts, latency and bandwidth, where latency is the fixed cost of each communication and bandwidth is the variable cost that depends on the amount of data to be transferred.
Latency in internode communications is a fundamental barrier to this goal, and advancements to network latency have historically been slower than improvements in other barriers to computing performance such as bandwidth and computational power~\cite{PattersonLatency}.
By avoiding external node communication until the domain of dependence has been exhausted, the calculation may advance many time steps with only a handful of communication events thereby improving the overall performance of the simulation.
This method is known as swept time-space decomposition~\cite{alhubail:16jcp, MalasHager}.

Recently, extreme-scale computing clusters have been used to perform CFD simulations on over 1.97 million CPU cores, each capable of four way hyper-threading.
This simulation was performed on a cluster with over \num{65000} \num{16}-core processors~\cite{BM_BigSolution}.
The monetary cost, power consumption, and size of such a cluster is a barrier to the realization of widespread peta- and exa-scale computing required for real-time, high-fidelity, CFD simulations.
While these are significant challenges, they also provide an opportunity to develop new tools that increase the utilization of the available hardware resources.
As the authors of CFD Vision 2030 note, ``High Performance Computing (HPC) hardware is progressing rapidly and is on the cusp of a paradigm shift in technology that may require a rethinking of current CFD algorithms and software''~\cite{slotnick:2014}.
Using Graphics Processing Unit (GPU) hardware as the primary computation device or an accelerator in a heterogeneous system is a viable, expedient option for cluster architecture that mitigates these problems.

GPU technology has developed rapidly recently progressing from Kepler to Pascal architecture in just four years, doubling peak single-precision performance and more than tripling peak double-precision performance~\cite{PascalWP:2016}.
The presence of GPUs in clusters, such as ONRL's Titan supercomputer, has become increasingly common in the last decade with the advent of NVIDIA GPU accelerators and Kepler architecture.
This has driven the development of software capable of utilizing and unifying the disparate architectures efficiently~\cite{Witherden20143028}.
The programs presented here are designed to be run on a single workstation rather than a compute cluster.
They are meant to study the swept rule as a method for arranging numerical computation with a view to accelerating simulations on workstations and providing a framework for effective implementations on heterogeneous computing systems.

The GPU's ability to accelerate computations comes from its thread and memory hierarchies discussed in section~\ref{GPUArch} and the CUDA language, a C-based API developed by NVIDIA for its GPUs.
These features are particularly well suited to solving PDEs in general and the swept rule in particular because they provide the programmer significant control over memory accesses and thread identification.
The swept rule, like other partitioning schemes such as cache-oblivious algorithms~\cite{Strzodka}, operate on a simple principle: do the most work possible on the values closest to the processor before fetching more values.
Because the data closest to the processor is also the most difficult to communicate, the strict application of this rule does not always provide the best performance, but it is a useful heuristic for implementing the procedure and analyzing its performance.

\section{Related work}

The work presented here~\cite{MyRepo} takes the results and ideas of Alhubail and Wang~\cite{alhubail:16jcp, MaithamRepo} as its starting point.
In their paper, they introduce the swept rule for explicit, time-stepping, numerical schemes applied to PDEs.
This method has elements in common with parallel-in-time and communication-avoiding algorithms.

Parallel-in-time methods, such as the multigrid-reduction-in-time algorithm~\cite{falgout2014parallel}, accelerate PDE solution methods by using time integrators to overcome the interdependence of solutions in the time domain allowing the entire space-time domain to be parallelized.
These methods calculate the solution over the space-time domain using a coarse grid and iterate over successively finer grids to achieve the desired accuracy.
The use of coarse grids in parallel-in-time methods causes a loss in efficiency and accuracy when applied to nonlinear systems~\cite{alhubail:16jcp}.
This shortcoming is intuitive; since chaotic, nonlinear systems may change suddenly in time, and coarse grids are prone to aliasing, the required grid granularity diminishes gains in performance.
The swept rule has the same motivation and application as parallel-in-time methods, but does not seek to parallelize the computation in time or vary the grid throughout the procedure.
It is therefore a more robust method for simulating chaotic systems.
The swept rule does not alter the numerical scheme; instead, it chooses where and when to evaluate the solution at a given point on the grid to compute the maximum number of solutions with the data stored in the memory most accessible to the processor.

In this way the swept rule shares many implementation details with communication-avoiding algorithms.
Recent developments in communication-avoiding algorithms involving GPUs have generally focused on applications involving matrices such as QR and LU factorization.
The LU factorization algorithm presented by Baboulin et al.~\cite{BABOULIN201217} is motivated by the increasing use of GPU accelerators in large-scale, heterogeneous clusters.
In this method, tasks are split between the GPU and CPU and communication between the devices is minimized.
This allows the communication and computation performed on each device to overlap so all data transfer is asynchronous with computation.
This approach, overlapping data transfer with hybrid computation, is explored in this paper.
Perhaps the recent development most applicable to this work is the communication-avoiding QR factorization algorithm presented by Anderson et al~\cite{Anderson}.
In their work the algorithm is designed for a single general-purpose GPU (GPGPU) in a desktop workstation and leverages well-known matrix manipulation algorithms.
They do not seek to alter or improve existing QR factorization algorithms; instead, they present methods for arranging and tuning existing algorithms for the best performance on a GPU.
Similarly, this paper focuses on the adaptation of the swept rule to the GPU using CUDA.
The swept rule is a strategy for arranging the computational path of explicit numerical methods, and this work seeks to tune the data structures and operations used in that path to achieve the best performance on the GPU.

\section{GPU architecture} \label{GPUArch}

The massively parallel processing capabilities of a modern GPU are enabled by a unique architecture originally designed to improve visual output from a computer.
The features of this architecture have been described in great detail by GPU producers and third-party authors~\cite{cudaProgGuide}~\cite{EngineerCuda}.  As particular aspects of this architecture, especially the unique and accessible memory hierarchy, are at the core of this research, some explanation of its relevant elements is necessary before describing the details of the implementation.

The Tesla K40c GPGPU has 15 streaming multiprocessors (SM), each capable of processing 64 warps of 32 threads or 2048 total threads at once.
Thread scheduling, where and when to carry out each thread's route through the computation, is controlled by the device hardware.
CUDA programs consist of functions, referred to as kernels, launched from a C\slash\CC{} host program.
The host code is executed on the CPU and specifies the size and number of blocks of threads that will operate on a kernel, the stream (queue) in which the kernel will be launched, and the amount of shared memory to be allocated per block at runtime.
All threads in a warp, a single-instruction group of threads, must be in the same block, so it is good practice to launch blocks with some multiple of 32 threads.
Several blocks may be in process simultaneously on an SM, and blocks may not be split between SMs.

While each SM is capable of handling 2048 resident threads, this number is usually significantly lower because each thread or block make demands on limited memory resources: shared memory and registers.
Each SM on the Tesla K40c has \SI{46}{\kilo\byte} of shared memory and 65536 registers available.
Registers are the most limited and quickest access type of memory.
Registers are only accessible to individual threads, but can be shared between threads in a warp with shuffle operations for devices with a compute capability of 3.5 or higher.
Shared memory is a portion of the L1 cache over which the programmer is given control and is only accessible to threads within a block.
That is, in order for a thread to read a value stored in shared memory in a different block, even if that block is on the same SM and active, the second block would need to write the value to global memory where the first thread can access it.
Global memory is the slowest to access and most plentiful memory type; it is where the data that is copied from or allocated within the host program is stored.
All variables passed to a kernel and large arrays declared therein are stored in global memory.

Other memory types in the CUDA memory hierarchy include constant, texture, and surface; of these, the work presented here only uses constant memory.
Constant memory is read-only, available to all kernels for the lifetime of an application, and quick to access when all threads access the same location.
This makes it a convenient and performance conscious choice for storing constant values of the governing equations that are calculated at runtime.

\section{Swept scheme implementation}
\subsection{Test cases}

Two test problems are used to demonstrate the performance and function of the swept rule in one dimension: the heat and Kuramoto--Sivashinsky equations.
The derivation of the procedures for the numerical solutions can be found in the~\nameref{App:AppA}.
The heat equation, Eq.~\eqref{eqn:HeatDisc}, was chosen because of its simplicity and familiarity.
It is applied to the problem of heat conduction in a one-dimensional rod with an initial temperature gradient and insulated ends:
\begin{equation}
\frac{\partial T}{\partial t} = \alpha \nabla^2 T \;,
\end{equation}
where $T$ is temperature, $t$ is time, and $\alpha$ is thermal diffusivity.
It is discretized with a second-order central difference scheme in space and a first-order forward difference in time:
\begin{equation}
T_i^{m+1} = \text{Fo} (T_{i+1}^m + T_{i-1}^m) + (1-2 \text{Fo}) T_i^m \;,
\label{eqn:HeatDisc}
\end{equation}
where $i$ is the spatial node index and $m$ is the time index corresponding to time $t^m$.
$\text{Fo}$ is the Fourier number defined as $ \text{Fo} = \frac{\alpha \Delta t}{\Delta x^2} $,
where $\Delta t$ is the timestep size and $\Delta x$ is the spatial grid size.

The Kuramoto--Sivashinsky (KS) equation was chosen to demonstrate the implementation of the
swept rule for higher-order, nonlinear PDEs difficult to solve accurately with
schemes that discretize rather than step through the time domain:
\begin{equation}
u_t = -(uu_x + u_{xx} + u_{xxxx})
    = -\left( \frac{1}{2} u_x^2 + u_{xx} + u_{xxxx} \right) \;,
\end{equation}
where $u$ is the dependent chaotic variable (e.g., fluid velocity).
It is discretized similarly to the heat equation, as shown in Eq.~\eqref{eqn:KSDisc},
with a finite difference central in space scheme, but decomposing the fourth spatial
derivative using central differencing requires a five-point stencil.
The chaotic nature of the problem necessitates a higher-order scheme in the time dimension;
therefore, an explicit, second-order Runge--Kutta scheme, also known as the midpoint method, is applied:
\begin{align}
\frac{u_i^{m+1}-u_i^m}{\Delta t}
    = -\left( \frac{(u_{i+1}^m)^2 - (u_{i-1}^m)^2}{4\Delta x} +
    \frac{u_{i+1}^m + u_{i-1}^m + 2u_i^m}{\Delta x^2} +
    \frac{u_{i+2}^m - 4u_{i+1}^m + 6u_i^m - 4u_{i-1}^m + u_{i-2}^m}{\Delta x^4} \right) \;.
    \label{eqn:KSDisc}
\end{align}
This problem uses a periodic initial and boundary conditions.
That is, the stencil at point $0$ includes points $n$ and $n-1$.

\subsection{First-order} \label{OrderOne}

\begin{figure}[!hbt]
	\centering
	\begin{subfigure}[t]{.67\textwidth}
		\centering
		\includegraphics[width=\textwidth]{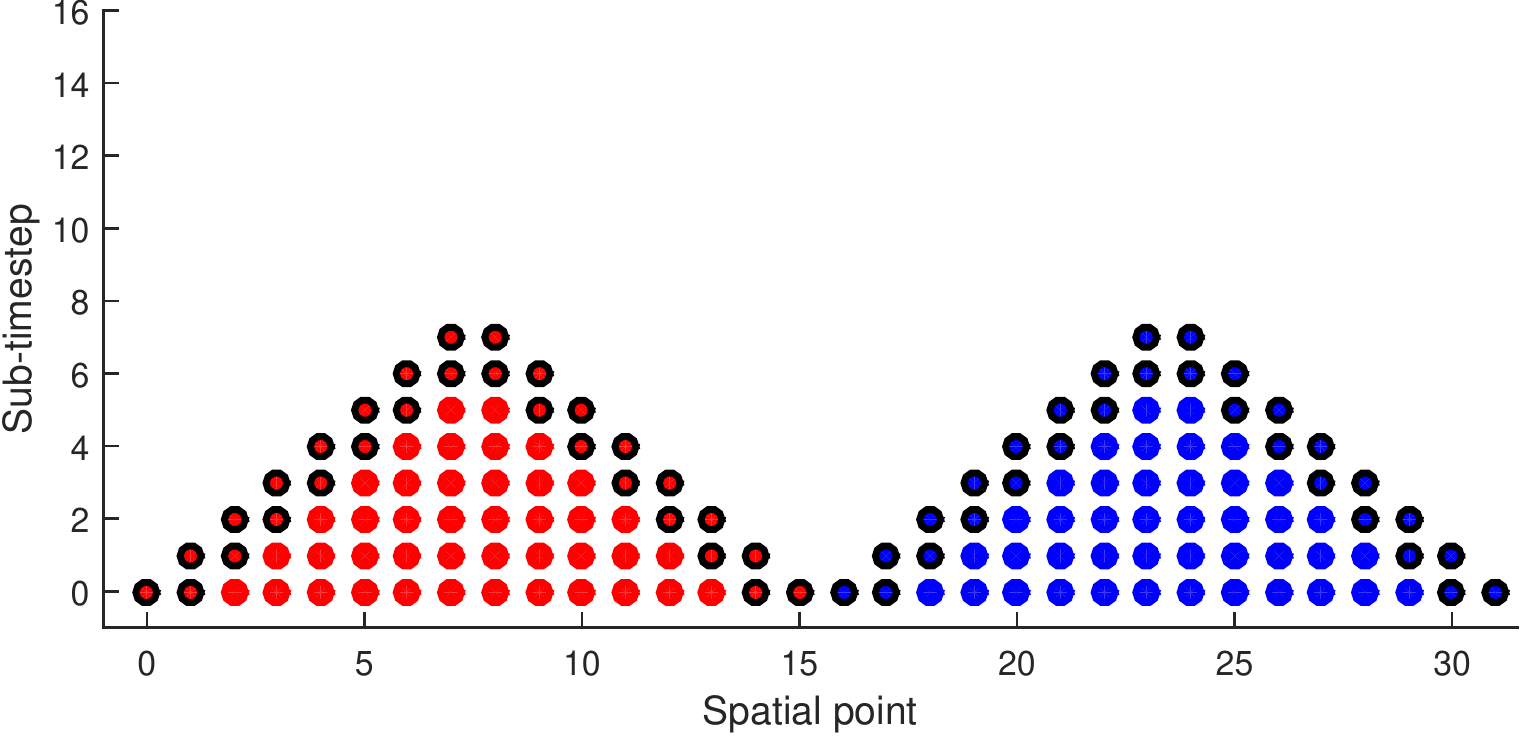}
		\caption{In the initial step, nodes 0 and 1, shown in red and blue respectively, compute a triangle in the space-time plane and collect the edge values in global arrays ``right'' and ``left'' shown with thick bordered dots.}
		\label{f:firstorder1}
	\end{subfigure}
	\hfill
	\\
	\begin{subfigure}[t]{.67\textwidth}
		\centering
		\includegraphics[width=\textwidth]{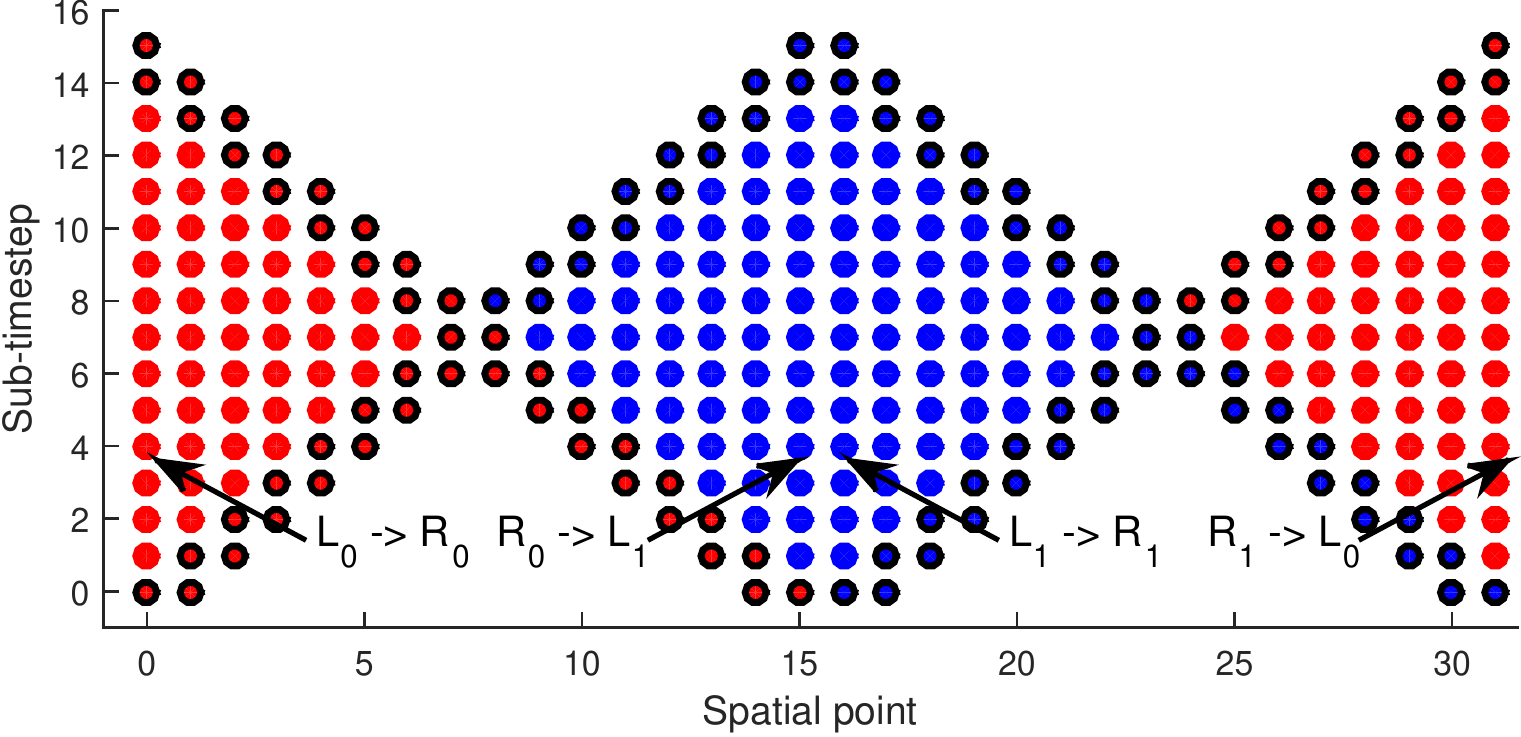}
		\caption{Next, the right edges are passed to the forward neighboring node and are swapped with the current node's left edge.
		Subsequently the opposite edges are swapped with the backward neighboring node.}
		\label{f:firstorder2}
	\end{subfigure}
	\caption{First-order scheme with three-point stencil.
	$L_i$/$R_i$ refer to left/right arrays of node $i$.}
	\label{f:firstorder}
\end{figure}

The initial incarnation of the swept rule presented by Alhubail and Wang~\cite{alhubail:16jcp,MaithamRepo} decomposes multi-step timesteps and large stencils into sub-timesteps with a three-point stencil.
This regularizes the procedure to accommodate various equations and methods, but increased the memory demands and number of iterations per timestep.
In the swept scheme's simplest form, applied to a linear, second-order PDE such as the 1D heat conduction equation with a first-order explicit method, each sub-timestep is a full timestep.
Figure~\ref{f:firstorder1} shows the progression of the computation cycle from the initial conditions in a node with $n = 16$ spatial points and $k = 2$ nodes.

When launching a CUDA kernel, it is strongly recommended that each block contain a multiple of 32 spatial points since there are 32 threads in a warp which execute in synchrony.
In view of the previously discussed GPU architecture and the overall aim of the swept rule, the most readily apparent execution model for the swept rule uses a block of threads as a node and shared memory.
Each thread is assigned a spatial point and advances in time until a member of its stencil is unknown.
The results are stored in a shared memory array so they are available to all threads within a block and can be retrieved as quickly as values from L1 cache as long as the access pattern is free of bank conflicts.

Shared memory values do not persist once a kernel has completed, and any data necessary to continue the computation must be shared between nodes.
So the information needed to begin a new nodal cycle must be read in from and out to global memory at the beginning and end of each kernel.
Since only the edges of each tier have unknown neighbor values, which are uncomputed or stored in another node's private memory, these values are ignored and only those dependent on local data are computed.
In one spatial dimension this forms a triangle of known values in the space-time plane whose two edge values on either side are necessary to form the three-point stencil for the complimentary triangle in the next nodal computation.
These carried values are stored separately in left and right global arrays and share the tip of the triangle.

\begin{figure}[!bh]
	\centering
	\begin{subfigure}[t]{.72\textwidth}
		\centering
		\includegraphics[width=\textwidth]{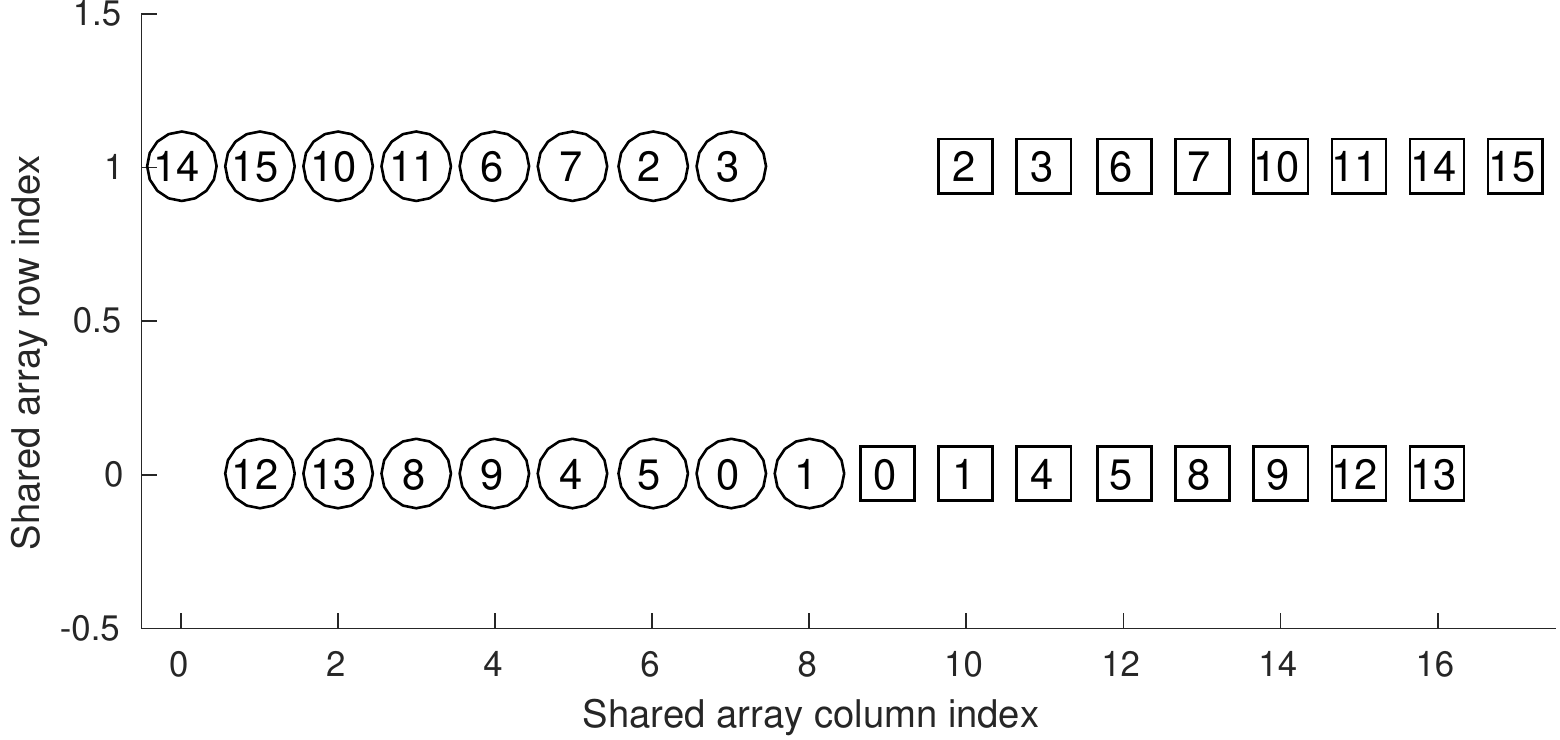}
		\caption{Edges of the locally computed triangle stored in shared memory are passed to their respective global arrays with the indices shown at the end of the kernel.}
		\label{f:insertion}
	\end{subfigure}
	\hfill
	\\
	\begin{subfigure}[b]{.72\textwidth}
		\includegraphics[width=\textwidth]{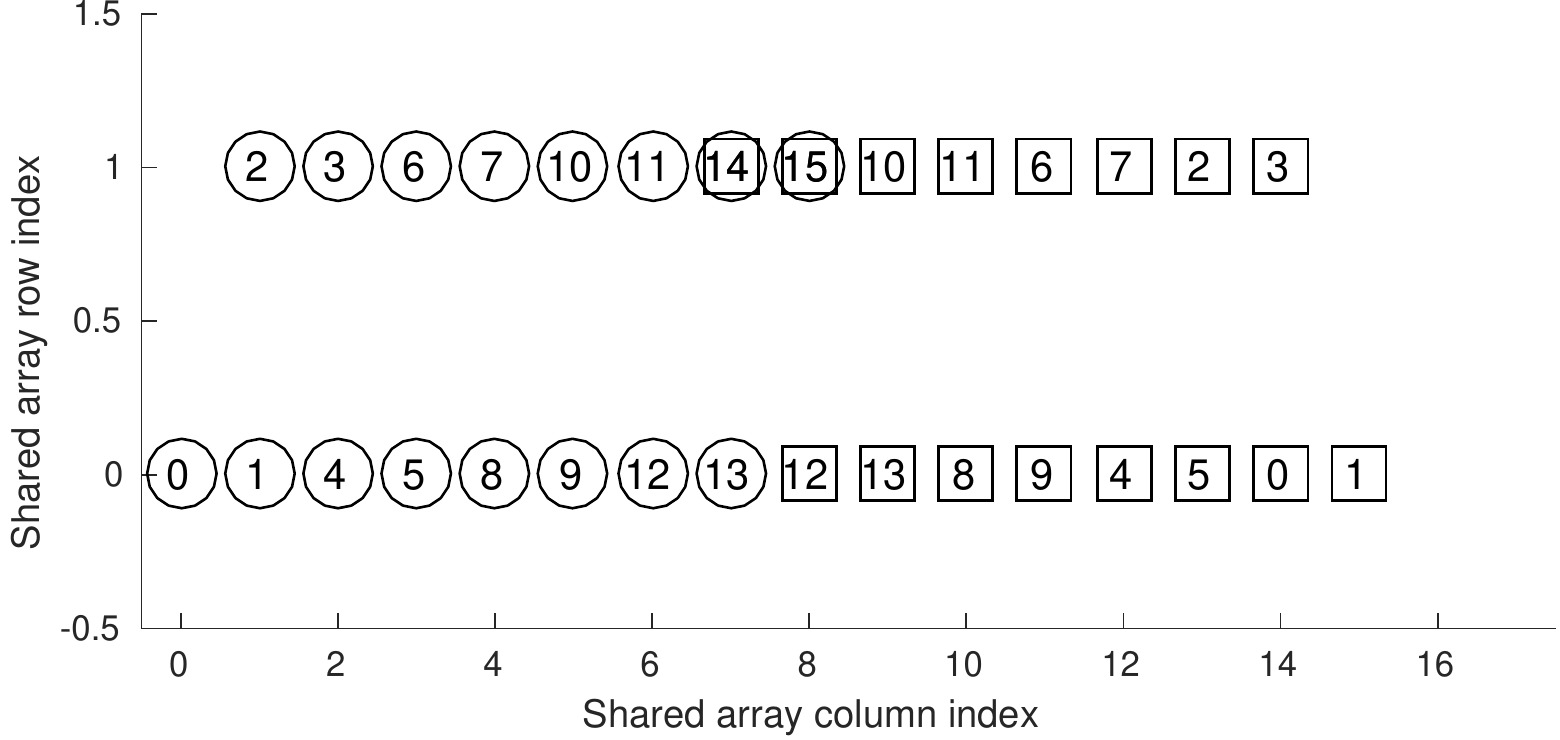}
		\caption{The edges are reinserted to seed the complimentary triangle after being swapped at the start of the kernel using the indices shown.}
		\label{f:extraction}
	\end{subfigure}
	\caption{The global arrays ``left'' and ``right'' represented by circles and squares, respectively.}
	\label{f:collection}
\end{figure}

Figure~\ref{f:firstorder2} shows why the two edges are stored individually: only one of the edges is passed between nodes.
After the first step, the right edge is passed to its right neighbor node; the left edge is stationary.
At the beginning of the next kernel the edges are swapped as they are reinserted as shown in Figure~\ref{f:insertion}.
For example, after the initial triangle is complete, node $1$ moves its right edge to node $k-1$ and receives the right edge of node $k-1$, and then swaps its left and right arrays.
When the right edge is passed between nodes, node $0$ is now split across the spatial boundary and must apply those boundary conditions at its center.

The swept rule allows the computation to advance by $n$ timesteps with two, rather than $n$, global memory accesses, where $n$ is the number of threads per block or number of spatial points per node.
The procedure advances further by passing values in the opposite direction and zig-zagging the location of the nodes in this fashion until the simulation is complete.
Since the diamonds, shown in Figure~\ref{f:firstorder2} do not store all the values at a single timestep, the simulation can only output values when a complimentary triangle is computed and the final $n$ length local tier is returned.
This kernel can only be called after the values are passed to the left, so the results can only be read out every $n_{th}$ timestep.

As described in section~\ref{GPUArch}, the number of resident threads in a kernel depends on the GPU architecture limitations and resources requested by the kernel launch.
For instance, to launch the program with $512$ threads per block with double-precision results and store every value in the triangle shown in Figure~\ref{f:firstorder1} in shared memory, each block would need to request $8 \times 65792 = \SI{514}{\kilo\byte}$ of shared memory, over $100$ times more than the \SI{46}{\kilo\byte} limit.
The maximum number of threads per block and blocks per SM would be limited to $128$ threads and $1$ block respectively.
This would have serious deleterious consequences on program performance, reducing parallelism by as much as an order of magnitude.

Figure~\ref{f:firstorder1} shows that the interior of the triangle is only needed to progress to the next timestep, and that the even and odd tier edges  do not overlap in the spatial domain.
Thus, the triangle may be stored as a matrix with two rows where the first row contains even sub-timestep values and the second contains odd values.
The interior values are overwritten once they are used, and the only values that remain are the edges.
Figure~\ref{f:extraction} shows the result of a local computation with this method.
The axes show the position of the edge values within the matrix, and the values on the figure refer to their index in the global left or right array.
The last two values, the tip of the triangle, are copied into both arrays.

\begin{figure}[!hb]
	\centering
	\begin{subfigure}[t]{.7\textwidth}
		\centering
		\includegraphics[width=\textwidth]{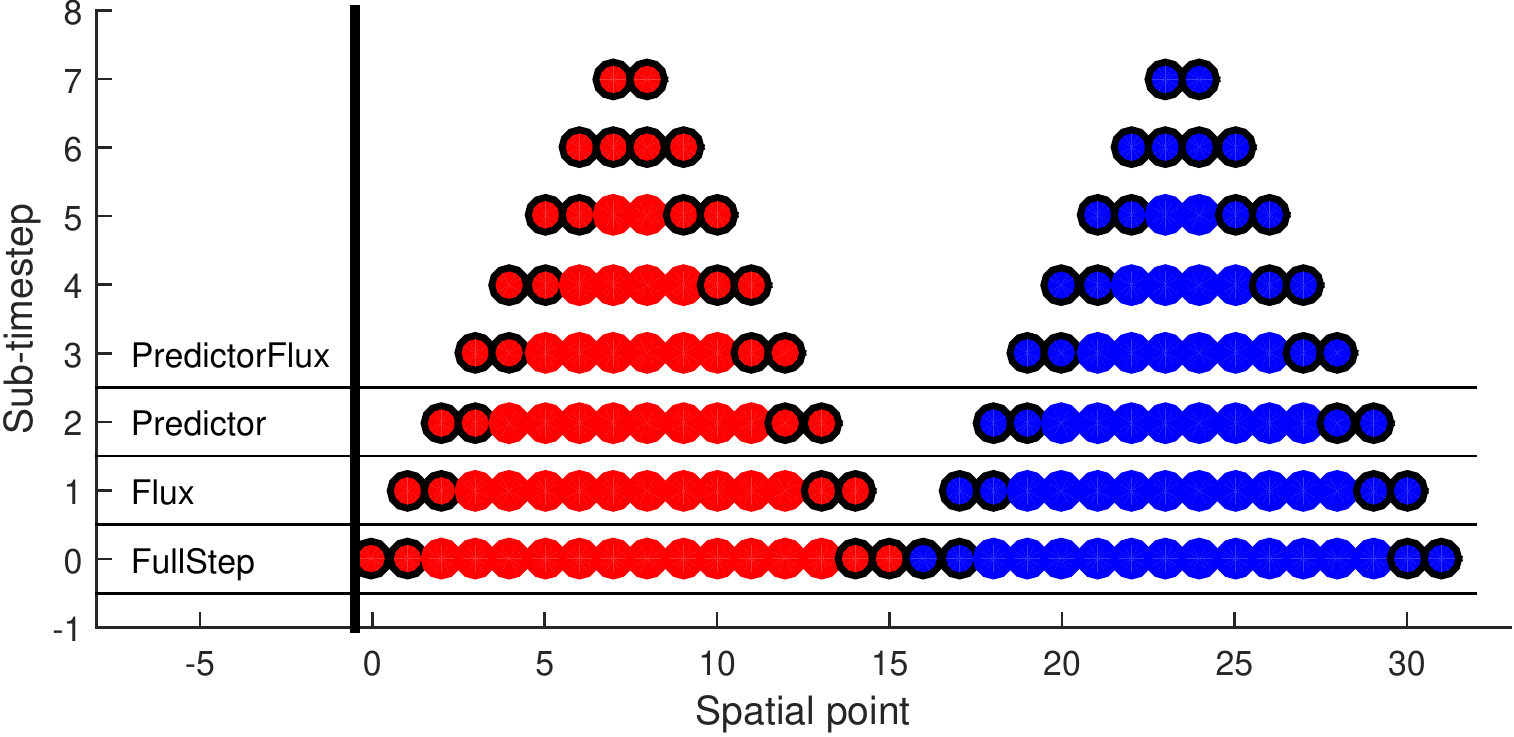}
		\caption{Second-order scheme with flux sub-timestep in the same form as first-order scheme.}
		\label{f:secondprob}
	\end{subfigure}
	\hfill
	\\
	\begin{subfigure}[b]{.7\textwidth}
		\includegraphics[width=\textwidth]{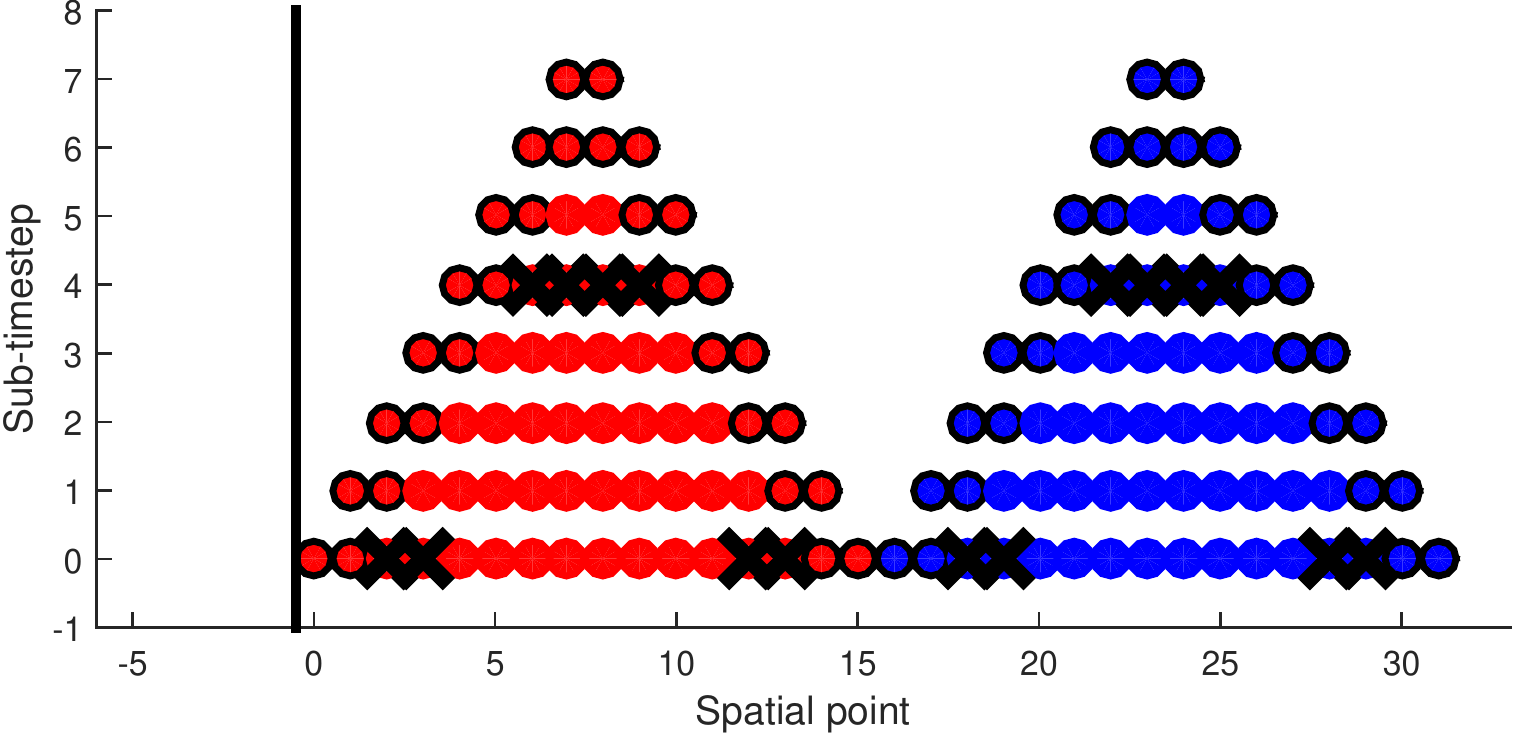}
		\caption{Values marked with ``x'' are overwritten before they are needed.}
		\label{f:secondexes}
	\end{subfigure}
	\caption{Splitting the procedure into sub-timesteps is a problem on a GPU with limited fast memory.}
	\label{f:secondorderProblem}
\end{figure}

To proceed in this fashion, the left and right arrays must be replaced to seed the solution matrix as shown in Figure~\ref{f:insertion}.
Now, the edges of the previous first row are moved to its center.
The top row's center, shared when the values were read out, is now left open for the first tier of the complimentary triangle solution.
The $x$ dimension is larger because two edge values are required on either side in order to complete the stencil for the longest row.
The computation proceeds by filling the empty two indices on the top row and then overwriting the bottom row's middle four indices and so on.
In contrast to the memory demands of storing the entire nodal computation, this method uses only $2\times(n+2)$ values.
For $n=512$, this requires $8 \times 1028 \approx \SI{8}{\kilo\byte}$.
Using only \SI{8}{\kilo\byte} means the kernel is not limited by shared memory usage and multiple blocks can be resident concurrently on each SM since these applications are not register limited.

\subsection{Second-order extended stencil} \label{OrderTwo}

In order to reduce the numerical scheme for the KS equation to a series of sub-timesteps requiring three-point stencils, the original swept scheme calculates $u_{xx}$ as a sub-timestep~\cite{alhubail:16jcp}.
Since this flux is required at neighboring spatial points to calculate the new timestep or predictor value, there are four sub-timesteps per timestep as shown in Figure~\ref{f:secondprob}.
This approach presents a storage and data transfer problem on the GPU.

Since the timestep value is used to calculate the next timestep, it must be saved to be used four tiers after it is computed.
In the previous method, several of these interior values would be overwritten on the next even tier sub-timestep.
These forgotten but required values are marked with $x$ in Figure~\ref{f:secondexes}.

\begin{figure}[!htb]
	\centering
	\begin{subfigure}[t]{.76\textwidth}
		\centering
		\includegraphics[width=\textwidth]{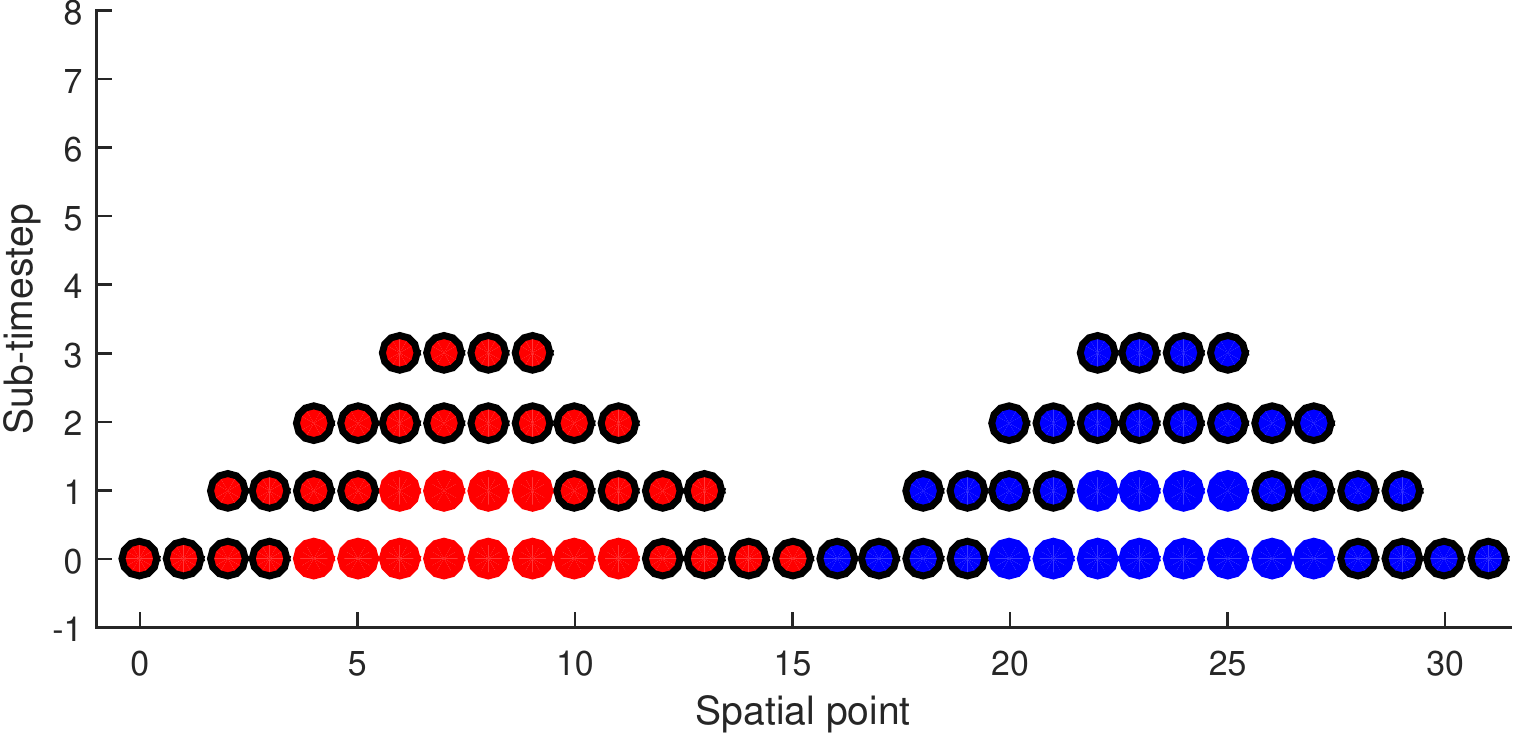}
		\caption{More values are required for stencils in this higher order scheme.  Thicker edges are passed from flatter triangles.}
		\label{f:secondsol1}
	\end{subfigure}
	\hfill
	\\
	\begin{subfigure}[b]{.76\textwidth}
		\centering
		\includegraphics[width=\textwidth]{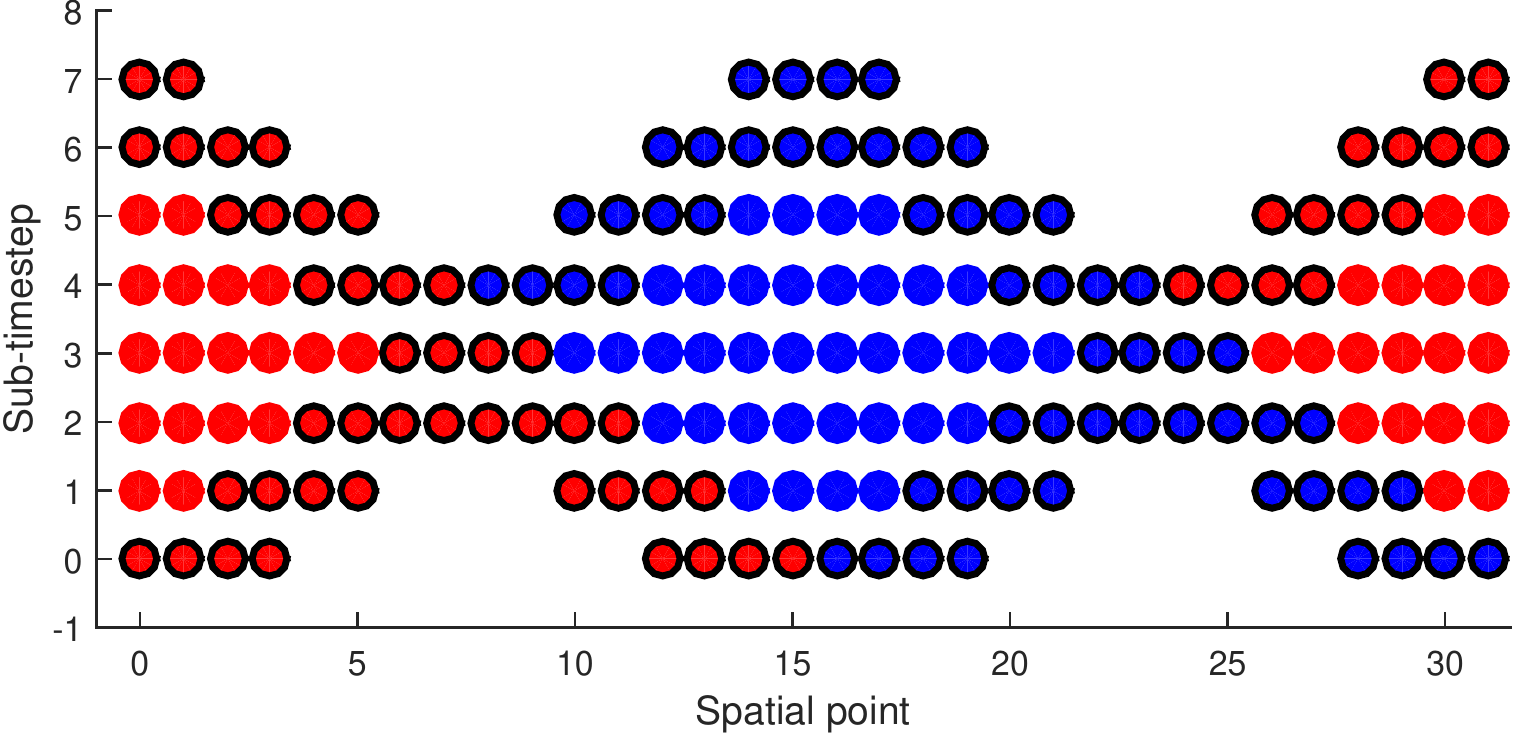}
		\caption{Values are passed between nodes as shown in Figure~\ref{f:firstorder2}.}
		\label{f:secondsol2}
	\end{subfigure}
	\caption{Flattened swept scheme includes flux calculation in sub-timestep.}
	\label{f:secondordersolution}
\end{figure}

Saving four values per tier would fix the problem, but would require a larger matrix in shared memory which would diminish occupancy, the amount of threads resident on an SM at a time, and require more unnecessary values to be communicated between nodes.
The solution shown in Figure~\ref{f:secondordersolution} includes the flux computation in each sub-timestep so a five-point stencil and two sub-timesteps per timestep---a predictor and a final value---are required.
This flattens the triangle or diamond and requires four values per sub-timestep, but the two row matrix may be used as described in Figure~\ref{f:collection}.
The same number of values are transferred between nodes in each communication, but inevitably more communications are required to advance the solution.
Conveniently, the predictor-corrector method ensures that all odd tiers, the second matrix row, will contain predictor values, and the bottom row will contain the final values.
The main loop of the kernel for the initial triangle is presented in Figure~\ref{f:SweptPseudo}.

\subsection{Classic scheme}

The classic kernel is a naive GPU implementation of the numerical solution and the baseline against which the efficacy of the swept rule is measured.
It advances one sub-timestep per kernel call and reads and writes solutions directly to global memory.
The structure of the procedure is exceedingly simple for the KS equation as shown in Figure~\ref{f:ClassicCode}.

\begin{figure}[htb]
\centering
\begin{minipage}[t]{0.55\textwidth}
\begin{lstlisting}
__global__ void
upTriangle(const REAL *IC, REAL *right, REAL *left) {
//tid = threadIdx.x (bottom row idx), tidT = tid+blockDim.x (top row idx);
//tids and tidTs: stencil for each tidT and tid respectively.
//Read initial data in from IC to shareT.

if (tid > 1 && tid <(blockDim.x-2)) shareT[tidT]=predictorStep(shareT, tids);
__syncthreads();

for (int k = 4; k<(blockDim.x/2); k+=4){

if (tid < (blockDim.x-k) && tid >= k) shareT[tid]+=finalStep(shareT, tidTs);

s2 = k + 2;
__syncthreads();

if(tid < (blockDim.x-s2) && tid >= step2) shareT[tidT]=predictorStep(shareT,tids);
__syncthreads(); }
//Lastly read out the edges to left and right arrays.
}
\end{lstlisting}
\caption{Main loop of the starting kernel for the swept rule.}
\label{f:SweptPseudo}
\end{minipage}
~
\begin{minipage}[t]{0.4\textwidth}
\begin{lstlisting}
__global__ void
classicKS(const REAL *ks_in, REAL *ks_out, bool final)
{
//Global Thread ID
int gid = blockDim.x * blockIdx.x + threadIdx.x;
//number of spatial points - 1
int lastidx = ((blockDim.x*gridDim.x)-1);
//Stencil indices.
int gids[5];

//True for all spatial points from periodic BCs.
#pragma unroll
for (int k = -2; k<3; k++)  gids[k+2] = (gid + k) & lastidx;

if (final) ks_out[gid] += finalStep(ks_in, gids);
else ks_out[gid] = predictorStep(ks_in, gids);
}
\end{lstlisting}
\caption{Classic kernel for KS solver. Final and predictor step functions can be written in C with \texttt{\_\_device\_\_} keyword.}
\label{f:ClassicCode}
\end{minipage}
\end{figure}

\subsection{Variants}

Though shared memory is the most obvious nodal storage solution for this algorithm, there are drawbacks to its use: high number of idle threads, poor utilization of CPU resources, and the fact that shared memory is not the fastest storage~\cite{harris:2014}.
Two alternative implementations are hybrid computation and register only computation.
Hybrid computation transfers a node between the CPU and GPU and performs the computation on the CPU.
This is a costly operation, but the devices can perform computations concurrently so if the CPU computation and transfers are less costly than the GPU kernel cycles, there is no penalty.
This strategy is applied to the heat equation because it eliminates the thread divergence involved in applying the boundary conditions in the middle of a node.
The first node, which would be split on the GPU, is passed to the CPU and each sub-timestep is executed in parallel with OpenMP.

Register-only computation is applied to the KS equation because it is easier to apply this procedure, which uses narrowly accessible memory, without enforcing boundary conditions.
In this implementation the number of points in a node is limited by the number of threads in a warp (32) which is constant over several iterations of NVIDIA GPUs.
Now, instead of reading the persistent data from global to shared memory and back, the values are read from global to shared to the registers and back.
Instead of the data residing in the memory location where the operations take place, the edges must be updated in the registers from shared memory after each function call and vice versa for the complimentary and regular triangles respectively.
This is because the shuffle operations that trade registers between threads only operate on active threads in the warp being called; if some threads are masked, they will be unable to supply the necessary stencil values.
Since they are included in the function call their values are overwritten and cannot be stored correctly at the start of the kernel as in the shared memory procedure.

\section{Results}

All tests presented here were performed on a single workstation with a Tesla K40c GPU and an Intel Xeon 2630-E5 CPU with eight cores and 16 potential threads~\cite{MyRepo}.
In the tests, the program was executed with all powers of two for \numrange{32}{1024} threads per block and $2^{11}$--$2^{20}$ spatial points.
Then, the best runs for each problem size were compared.
The solutions are verified by comparing the numerical to the exact solution for the heat equation and comparing the classic decomposition to the swept rule solution for the KS equation since there is no exact solution for periodic boundary conditions.

\begin{figure}[!htb]
	\centering
	\includegraphics[width=\textwidth]{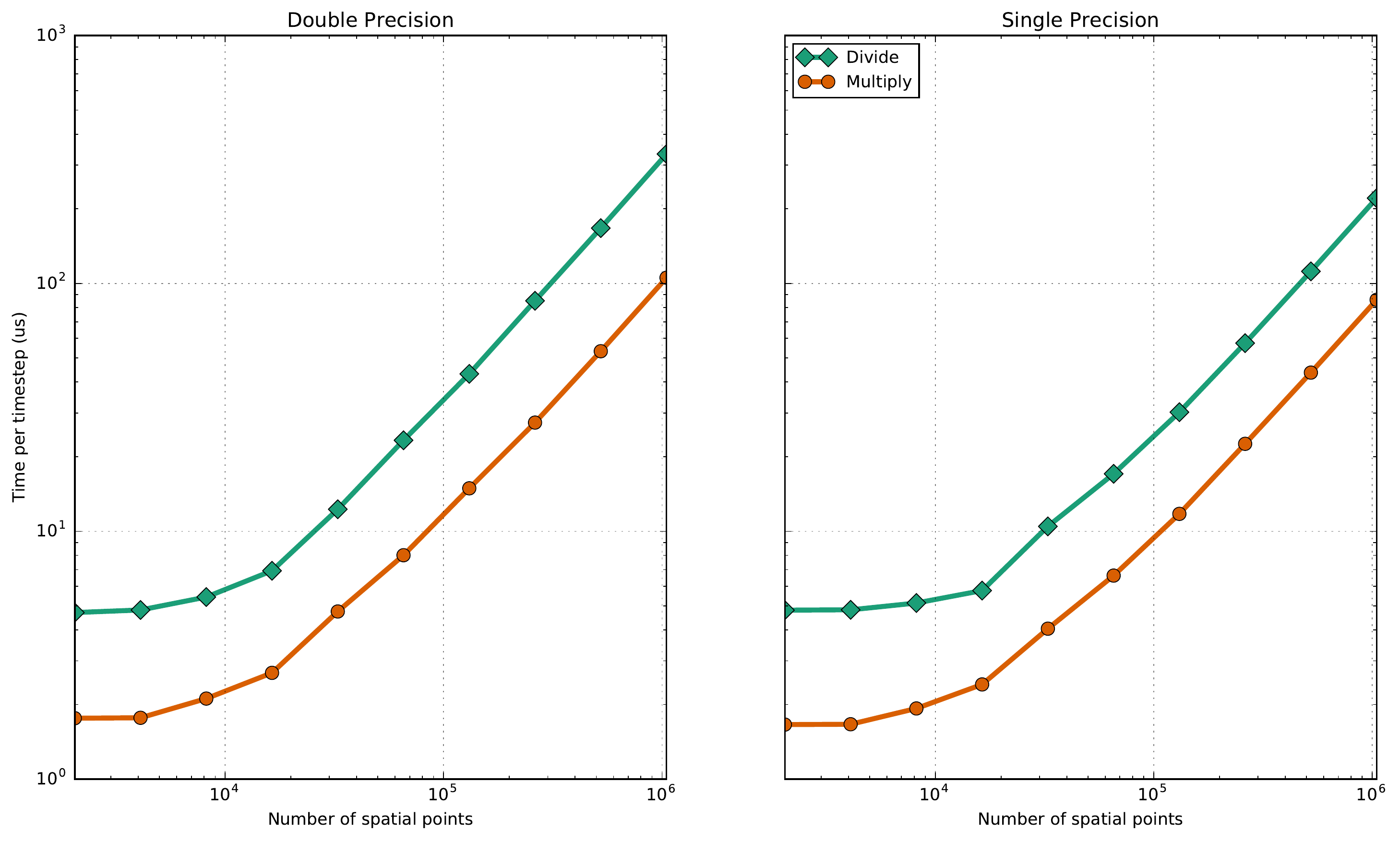}
	\caption{Performance effect of avoiding division operations.}
	\label{f:divisionComparison}
\end{figure}

In addition to the conceptually distinct implementations investigated herein, it bears presenting a general feature of the program that resulted in significant performance gains.
In these programs, constant values in the discretizations such as the Fourier number or $\Delta x$ are stored in a structure in constant memory on the device.
This is both efficient and convenient since the best constant memory performance is achieved when the value read from constant memory is broadcast; i.e., each thread in a warp reads the same value.
In general, mathematical representations of numerical methods construct the discretization by dividing by a gridstep or timestep.
In reality a program could just as easily store the inverse of the gridstep and multiply rather than divide during the computation.
Multiplication is a much cheaper instruction and its impact on the performance of the program is shown in Figure~\ref{f:divisionComparison}.
Avoiding division in just these instructions, the three denominators shown in Eq.~\eqref{eqn:KSDisc}, produces a speedup of about \SI{3}{$\times$} for the best block size for all problem sizes and precisions.

\begin{figure}[!htb]
	\centering
	\begin{subfigure}[t]{.48\textwidth}
		\centering
		\includegraphics[width=\textwidth]{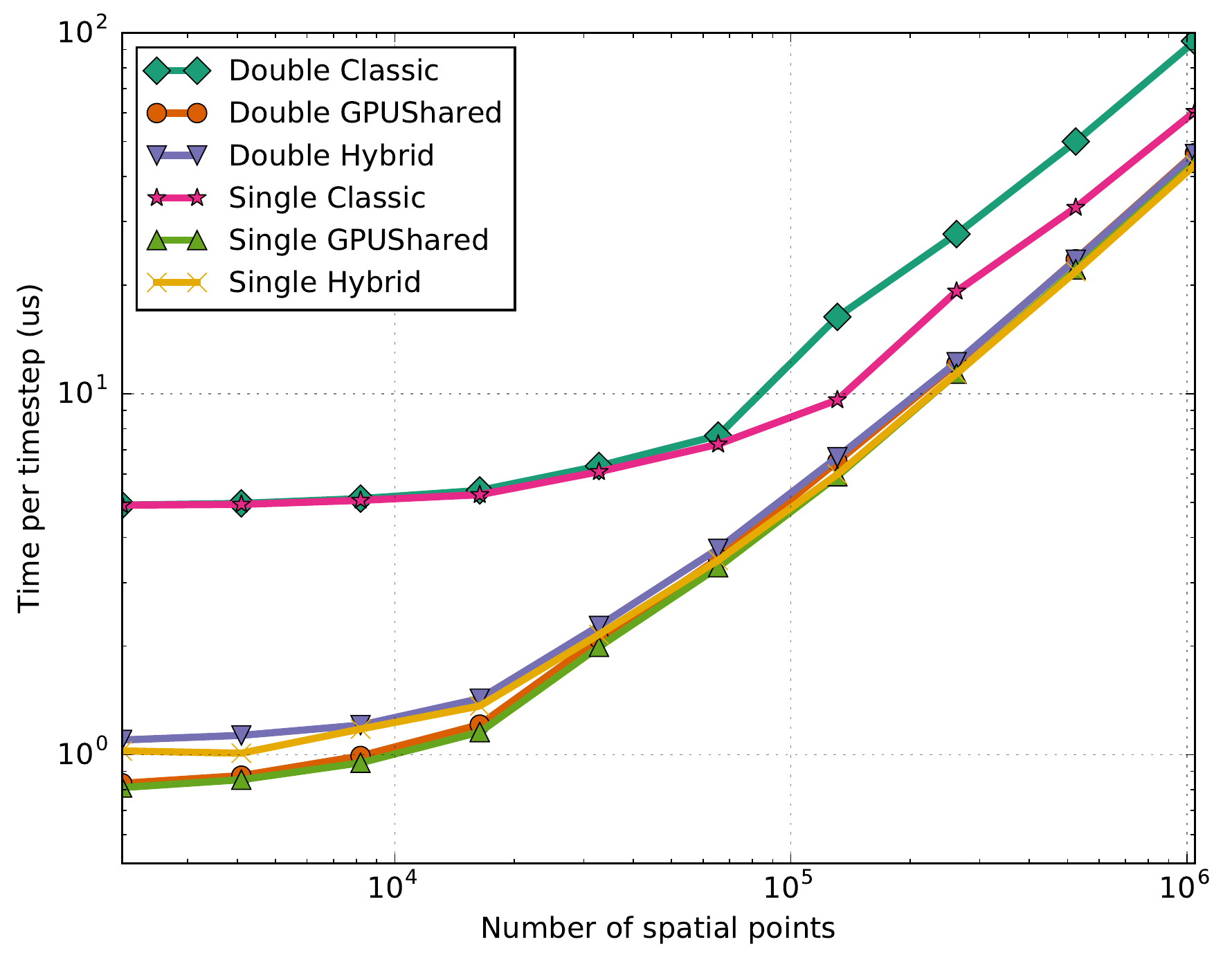}
		\caption{Performance of implementation concepts with best performing node size at each problem size and precision.}
		\label{f:HeatRaw}
	\end{subfigure}
	\hfill
	\begin{subfigure}[t]{.48\textwidth}
		\centering
		\includegraphics[width=\textwidth]{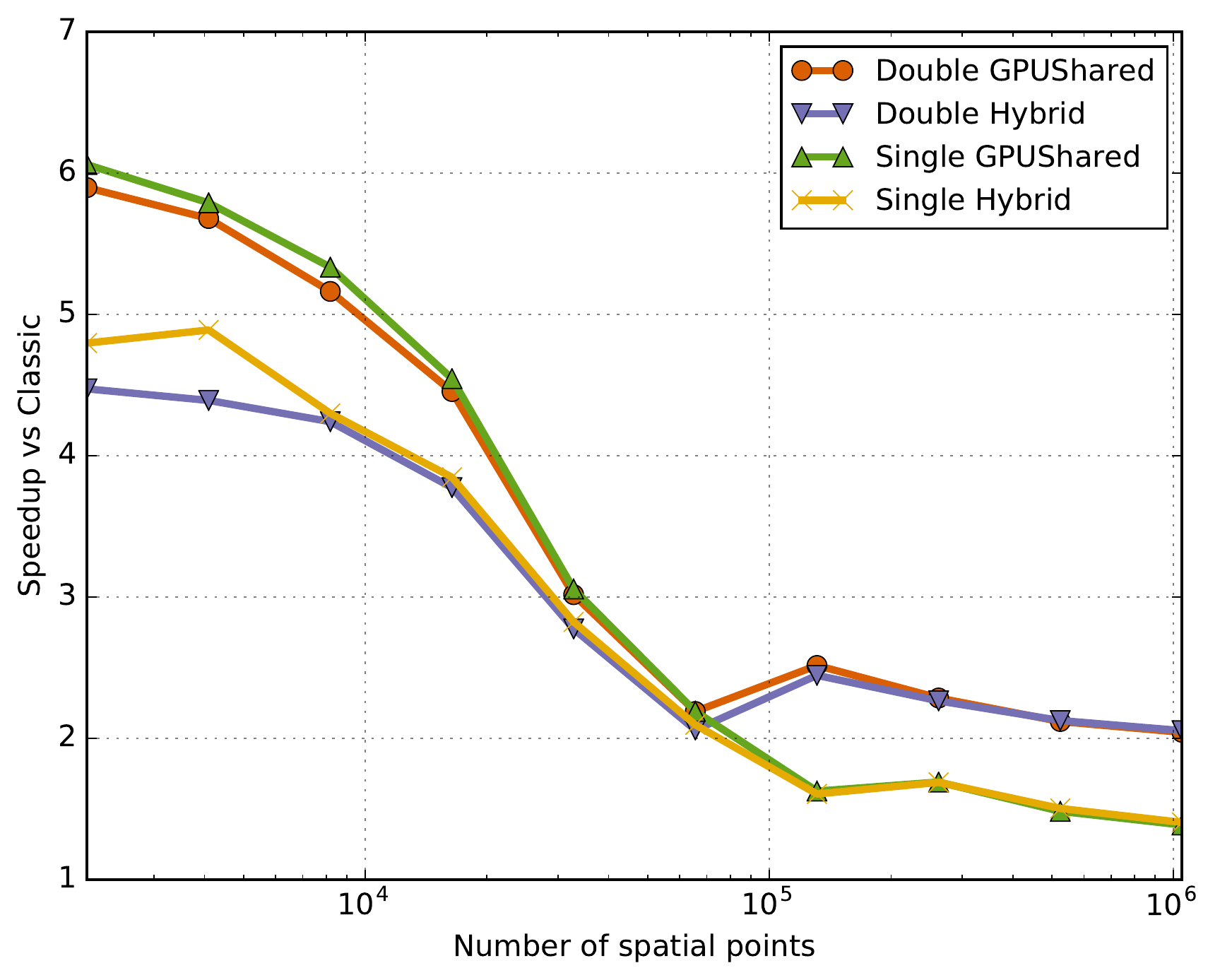}
		\caption{Speedup of swept rule programs with respect to the classic version at each problem size.}
		\label{f:HeatSpeed}
	\end{subfigure}
	\caption{Results of the heat equation test.}
	\label{f:heatResult}
\end{figure}

\begin{figure}[!htb]
	\centering
	\begin{subfigure}[t]{.48\textwidth}
		\centering
		\includegraphics[width=\textwidth]{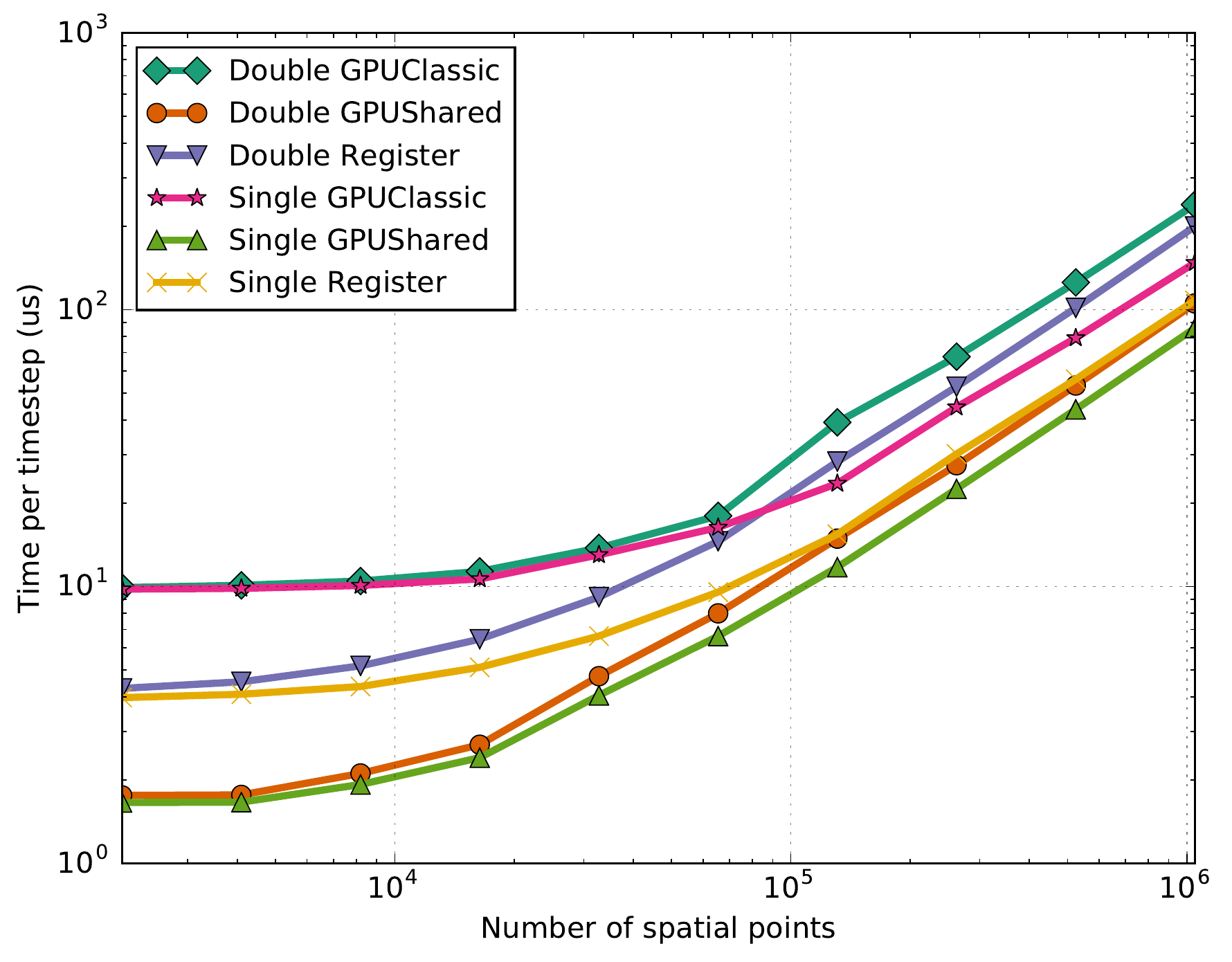}
		\caption{Performance of implementation concepts with best performing node size at each problem size and precision.}
		\label{f:KSRaw}
	\end{subfigure}
	\hfill
	\begin{subfigure}[t]{.48\textwidth}
		\centering
		\includegraphics[width=\textwidth]{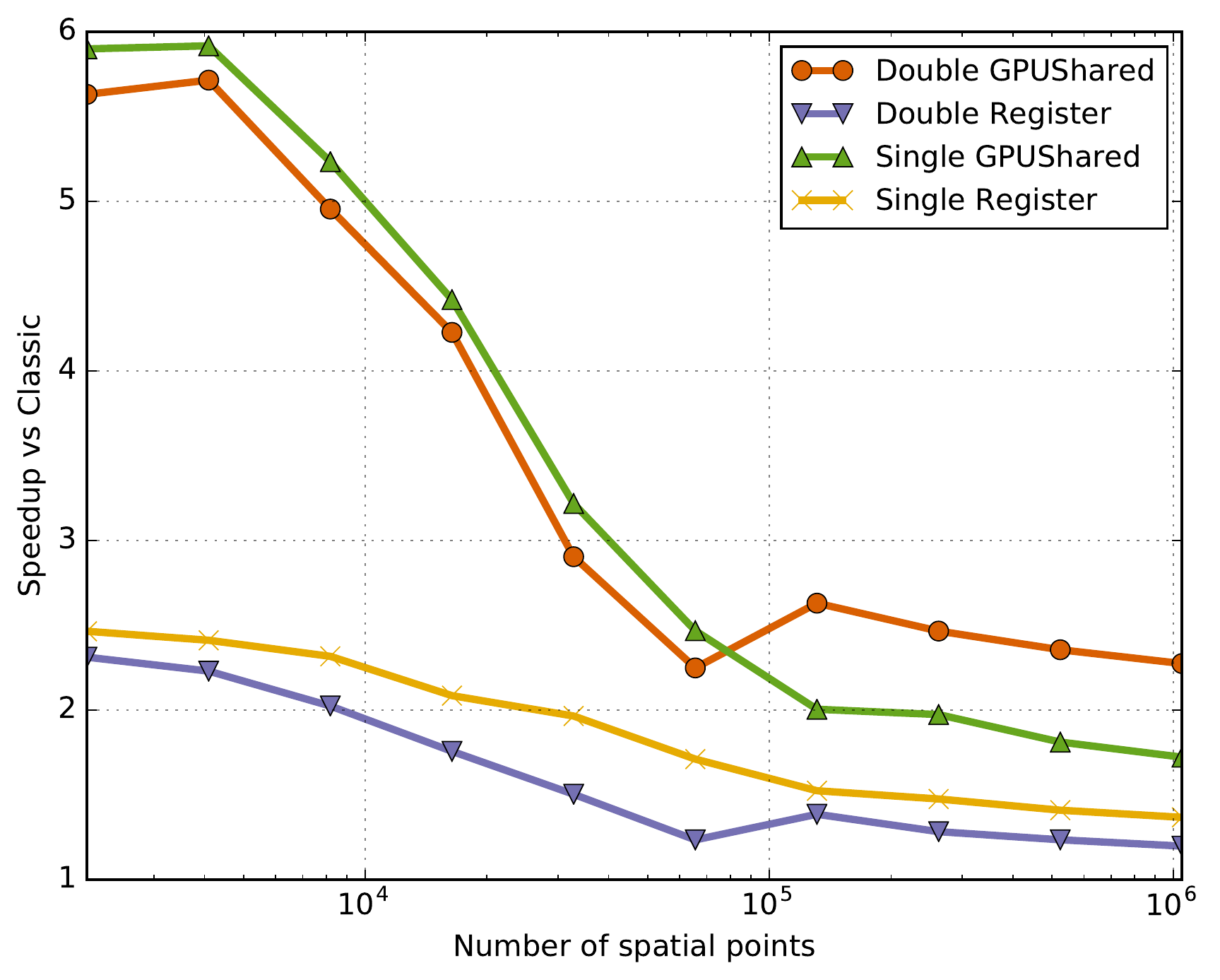}
		\caption{Speedup of swept rule programs with respect to the classic version at each problem size.}
		\label{f:KSSpeed}
	\end{subfigure}
	\caption{Results of the Kuramoto--Sivashinsky equation test.}
	\label{f:KSResult}
\end{figure}

Figures~\ref{f:HeatRaw} and~\ref{f:KSRaw} show the raw execution time per timestep of the GPU solutions and the performance of the swept rule with respect to the naive solution.
In all cases the swept rule outperforms the classic procedure (\texttt{GPUClassic}), and the GPU-only version based in shared memory, \texttt{GPUShared}, beats the alternate swept rule program.
The speedup of the swept rule, shown in Figures~\ref{f:HeatSpeed} and~\ref{f:KSSpeed}, is the ratio of its execution time to the classic version at the same precision and problem size.
Both cases have similar performance patterns; \texttt{GPUShared} generally provides a \SI{6}{$\times$} speedup for small problem sizes, but only a \SI{2}{$\times$} speedup for large problem sizes.
Figures~\ref{f:HeatRaw} and \ref{f:KSRaw} show that as problem size becomes large the execution time of all the programs grows at a similar rate.
In addition, all programs have an initial range where the execution time is relatively insensitive to problem size.
However, the swept rule versions begin to grow in cost at smaller problem sizes.

\begin{figure}[htb]
	\centering
	\begin{subfigure}[t]{.48\textwidth}
		\centering
		\includegraphics[width=\textwidth]{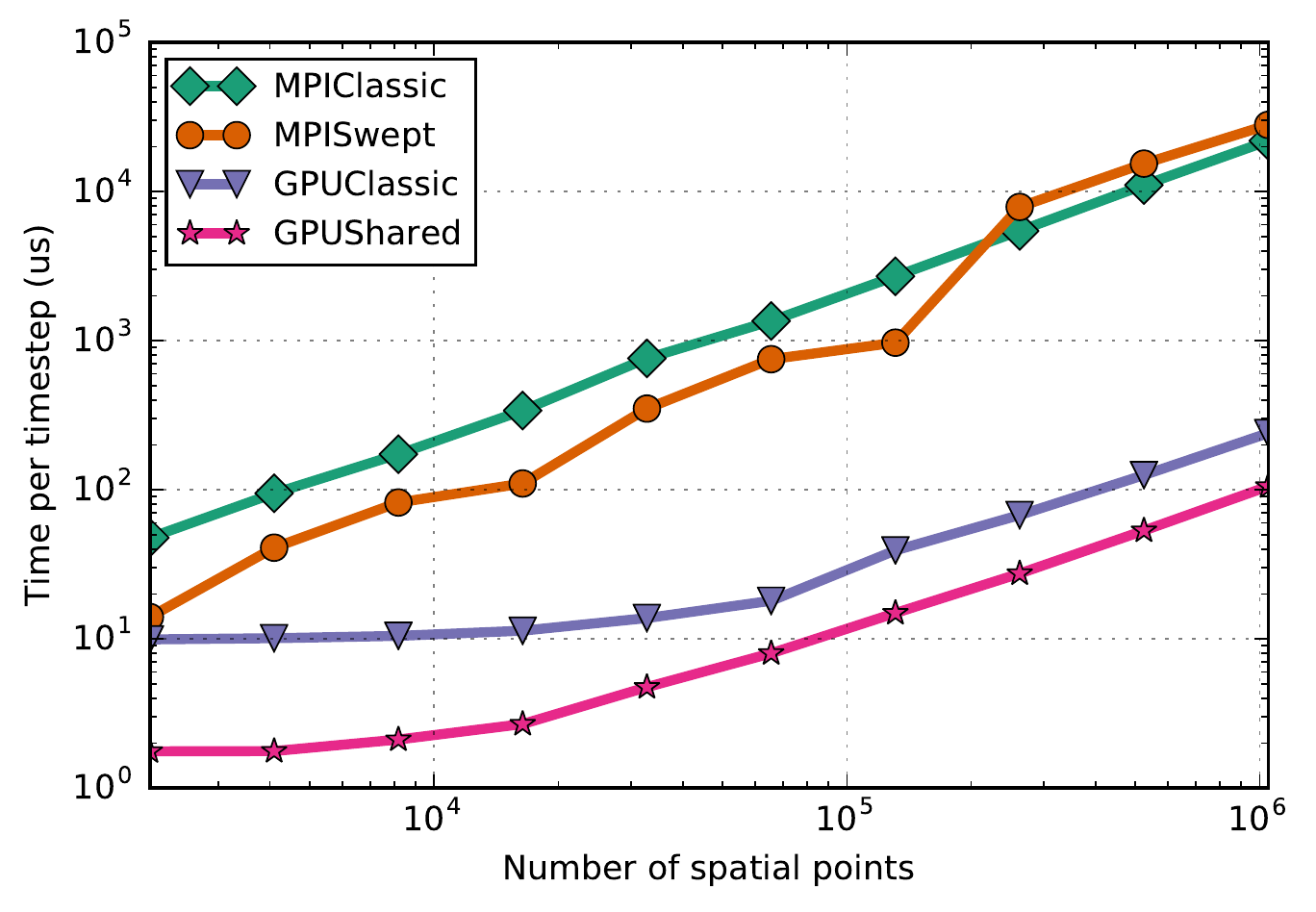}
		\caption{Performance test results of original swept rule MPI and CUDA programs for KS equation.}
		\label{f:mpiraw}
	\end{subfigure}
	\hfill
	\begin{subfigure}[t]{.48\textwidth}
		\centering
		\includegraphics[width=\textwidth]{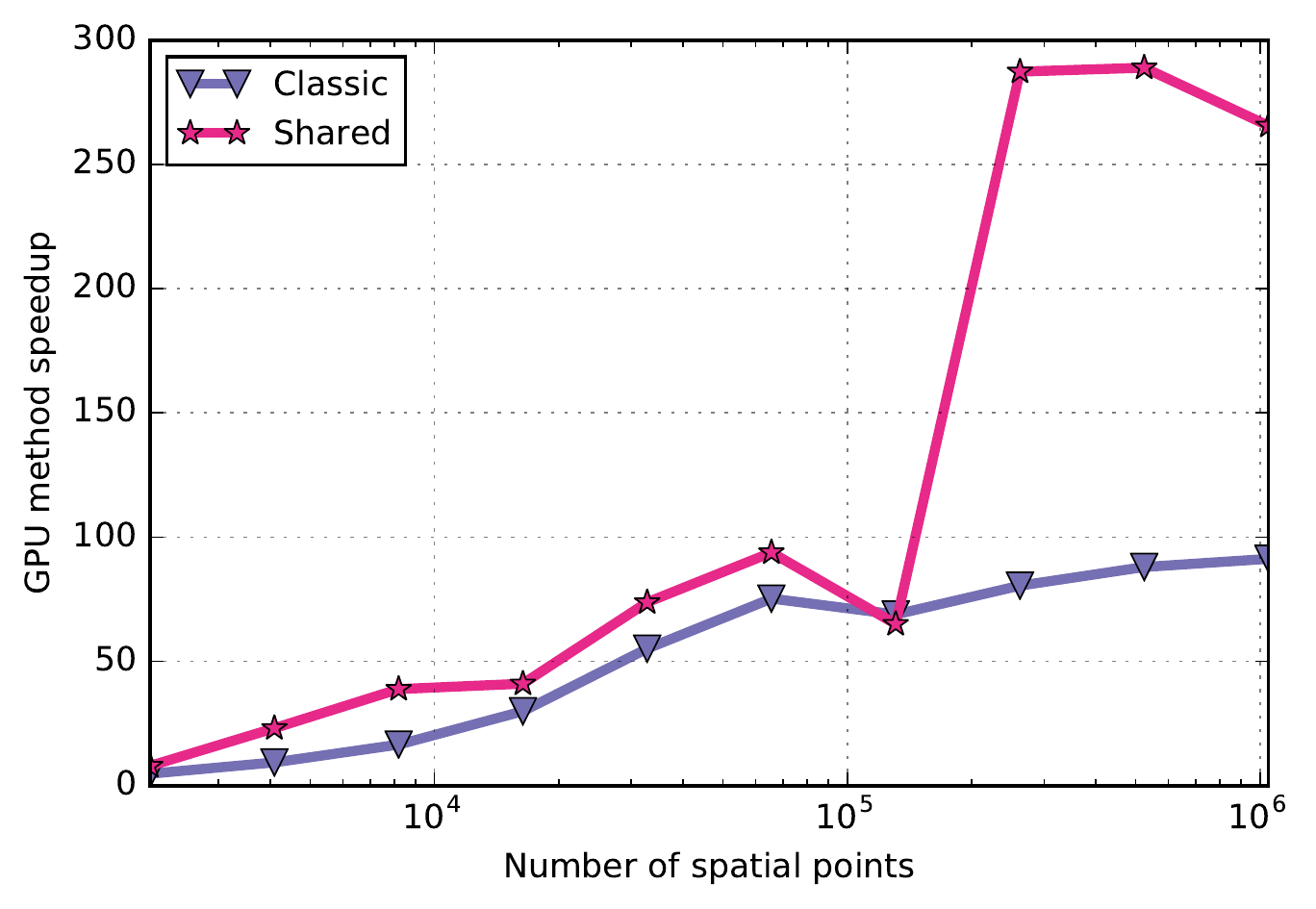}
		\caption{Performance improvement of GPU program vs. analogous MPI version.}
		\label{f:mpispeed}
	\end{subfigure}
	\caption{KS CPU version using MPI compared to GPU versions.}
	\label{f:mpi}
\end{figure}

These results are similar to the findings of the original swept rule case study~\cite{alhubail:16jcp}, which compared the swept and classic decompositions and found their execution times to be the same at large node sizes.
Because of the differences in GPU architecture, nodes cannot be arbitrarily large so the best node size for each problem size and precision is compared.
The original swept rule uses the message passing interface (MPI) library to implement the procedure so it may be evaluated in parallel by a single processor or a cluster.
This results in performance improvements for the KS equation with node sizes from 10 to \num{10000} spatial points, with the most significant improvements of about \SI{25}{$\times$}  occurring at about 100 spatial points.
The results of the MPI-based swept rule program~\cite{MaithamRepo} for the KS equation are shown in Figure~\ref{f:mpiraw} alongside the double-precision results also found in Figure~\ref{f:KSRaw}.

It should be noted that the original scheme is designed with a computational cluster in mind, and that the performance study presented by Alhubail and Wang~\cite{alhubail:16jcp, MaithamRepo} was executed with only two threads on two CPUs in a cluster.
The MPI program was tested in a similar way as the GPU program for this project; it was run on the workstation CPU with various problem sizes and number of threads up the maximum (i.e., 16).
The best result for each problem size, usually with 16 threads, is compared.
This did not degrade the performance compared to the original study; in fact, both versions of the program performed significantly better, in most cases the swept program performed about \SI{6}{$\times$} better.
Since the test used up to eight times the number of threads, this is a reasonable result that supports the validity of repurposing this code and comparing the result to the GPU versions.
The GPU programs are always faster than their MPI counterparts in the tested range.
The improvement of the GPU versions increases with problem size which is the opposite of the trend of the swept rule's improvement with respect to the classic kernel.
Both discretization types see a speedup of about \SI{5}{$\times$} for the smallest problem sizes and the speedup is about \SIrange{300}{100}{$\times$} for the swept and classic discretizations, respectively, at large problem sizes.

\section{Conclusion}

The decrease in the performance improvement of the swept rule compared to the classic discretization is mainly caused by occupancy.
This affects performance in two ways; first, the swept rule uses more shared memory, registers, and control flow than the classic version.
These resource requests diminish the number of threads that can be resident on an SM, so the kernel launches waves of blocks on each SM.
Each wave must wait until the previous one is complete before beginning so theoretically two waves will cost twice as much as one.
As more blocks are launched for larger problem sizes the difference in the number of waves per kernel call increases and diminishes the advantage of the swept rule.
Second, the classic kernel depends heavily on occupancy for its performance. GPU memory accesses rely on hiding latency with occupancy by switching between warps while others are waiting for dependent values.
If the GPU utilization is well below capacity, for example launching eight blocks of 256 threads for problem size \num{2048}, then only eight of the SMs are active and they contain only eight out of a possible 64 warps.
The swept rule also relies on occupancy for performance, since utilizing as many available resources as possible is always good for performance, but less than classic since global memory accesses are a smaller part of the procedure.

Figure~\ref{f:mpi} is more illustrative of the effect of GPU architecture than the swept rule.
At small problem sizes, under \num{e4}, the GPU is capable of assigning one thread to each spatial point and processing them concurrently, although not simultaneously.
There its improvement over the MPI version of the schemes is modest, under 50 times, for the same reasons that the swept rule's speedup diminishes with increasing problem size: the GPU is not given enough work.
With larger problems that fully occupy the GPU, the performance increases continue growing.

All figures in the previous section show little influence of precision on execution time and a penalty for greater precision emerges only at larger problem sizes.
The precision penalty is insignificant at lower problem sizes where the GPU is not fully occupied, but emerges at the point where multiple waves of blocks are launched.
These particular programs have relatively low register requirements and do not exhaust shared memory supply which leads to high occupancy for both precisions, but the small difference in register usage from double-precision values lead to a difference in theoretical occupancy which causes the difference in the number of waves required for each kernel launch to grow.
This conclusion implies a prediction: more complex numerical schemes and problems that require storing multiple dependent variables, i.e., systems of equations, will exhibit a larger effect of precision on performance.

Both alternate swept-rule procedures, hybrid processing heat and register swapping, performed worse than the \texttt{GPUShared} version.
The failure of the hybrid processing routine has less to do with the cost of host-to-device communication, since that is handled with asynchronous communication using streams, than improvements made to the shared memory program's handling of the boundary conditions.
Assigning a thread private (in register) array of stencil indices at the start of the kernel applies the boundary conditions with one instance of thread divergent control flow in the kernel as opposed to every sub-timestep.
The hybrid computation seeks to solve a problem that is more efficient to solve within the GPU paradigm; the register only computation shows more promise, but falls short because the node size is limited.
Since a node is the size of a warp, it must be 32 spatial points which results infour timesteps per cycle for the KS equation, which only reduces instances of communication by $1/8$.
Despite this limitation, its execution time increases more slowly with problem size at small and medium problem sizes.
Improvements to the edge swapping routines described in Figure~\ref{f:firstorder} could provide a significant benefit to the register based kernel making it competitive with the shared memory based implementation.

Perhaps the most important general lesson from this study is that the purely CUDA portion of a classic style implementation of an explicit numerical scheme can be written in 10 lines and result in multiple orders of magnitude performance improvement.
For most applications this is enough, but in cases where every morsel of performance is critical, the swept rule can further improve execution time.
This performance improvement is, of course, dependent on the hardware.
A commercial GeForce GPU with Kepler architecture would produce worse performance than the Tesla K40c which is designed for computation as opposed to actual graphics.
Despite the significant performance increase shown here, the Tesla K40c is three-year-old technology.
The current state-of-the-art Pascal architecture GPGPU, the Tesla P100, has twice the K40c base clock speed, nearly four times the number of streaming multiprocessors, and an extra \SI{16}{\kilo\byte} of shared memory.
It would be reasonable to predict that this device would halve the execution times shown here and maintain insensitivity to problem size up to over \num{e5} spatial points resulting in speedups on the order of \num{e3} over CPU parallel versions.

Future work will focus on characterizing the performance of the swept rule on heterogenous cluster architectures using CUDA-aware MPI and implementing the 2D swept rule on the GPU.

\section*{Appendix}
\label{App:AppA}

Here the derivations of the discretizations for the heat and Kuramoto--Sivashinsky equations are provided.
The heat equation without volumetric heat flux is
\begin{equation}
\frac{\partial T}{\partial t} = \alpha \nabla^2 T \;.
\end{equation}
In one dimension, this takes the form
\begin{equation}
\frac{\partial T}{\partial t} = \alpha \frac{\partial^2 T}{\partial x^2} \;.
\label{eq:heat1d}
\end{equation}
Discretizing Eq.~\ref{eq:heat1d} with forward differencing in time and central differencing in space:
\begin{equation}
\frac{T_i^{m+1}-T_i^m}{\Delta t} = \alpha \frac{T_{i+1}^m + T_{i-1}^m + 2T_i^m}{\Delta x^2}
\end{equation}
Defining the Fourier number $\text{Fo} = \frac{\alpha \Delta t}{\Delta x^2} $, and solving for temperature at the next timestep $T_i^{m+1}$:
\begin{equation}
T_i^{m+1} = \text{Fo} (T_{i+1}^m + T_{i-1}^m) + (1 - 2 \text{Fo}) T_i^m \;.
\end{equation}
This is a first-order, explicit, finite-difference method.
With insulated boundary conditions at both ends and $n$ spatial points:
\begin{equation}
T_0 = T_1 \quad \text{and} \quad T_{n+1} = T_{n-1} \;.
\end{equation}

The Kuramoto--Sivashinsky equation is a nonlinear, fourth-order, one-dimensional partial differential equation:
\begin{equation}
u_t = -(uu_x + u_{xx} + u_{xxxx})
= -\left( \frac{1}{2} u_x^2 + u_{xx} + u_{xxxx} \right) \;.
\end{equation}

Discretizing the fourth derivative requires neighbors of neighbors in the spatial dimension or an additional step where $u_{xx}$ is calculated for the three points in the domain.
Finite difference discretization with neighbors of neighbors:
\begin{equation}
\frac{u_i^{m+1}-u_i^m}{\Delta t}
= -\left(\frac{(u_{i+1}^m)^2 - (u_{i-1}^m)^2}{4\Delta x} +
\frac{u_{i+1}^m + u_{i-1}^m + 2u_i^m}{\Delta x^2}+
\frac{u_{i+2}^m - 4u_{i+1}^m + 6u_i^m - 4u_{i-1}^m + u_{i-2}^m}{\Delta x^4}\right) \;.
\end{equation}
Finite difference discretization with flux step:
\begin{align}
u_{xx} + u_{xxxx} &= \frac{\partial^2 u}{\partial x^2} + \frac{\partial^4 u}{\partial x^4}
=\frac{\partial^2}{\partial x^2}u + \frac{\partial^2}{\partial x^2}u_{xx}
= \frac{\partial^2}{\partial x^2} \left(u+u_{xx}\right) \\
(u_{xx})_i &= \frac{u_{i-1} + u_{i+1} - 2u_i}{\Delta x^2} \text{ at } i-1, i, i+1 \\
\frac{u_i^{m+1}-u_i^m}{\Delta t}
&= -\left(\frac{(u_{i+1}^m)^2 - (u_{i-1}^m)^2}{4\Delta x} +
\frac{(u+u_{xx})_{i+1}^m + (u+u_{xx})_{i-1}^m + 2(u+u_{xx})_i^m}{\Delta x^2}\right) \;.
\end{align}

Let the right-hand side be $f(u(x,t))$. Using a second-order, explicit Runge--Kutta method the solution at the new timestep is calculated in two steps.
First, the predictor solution is obtained at $u_i^{m+1/2}$:
\begin{equation}
u_i^{m+1/2} = u_i^m + \frac{\Delta t}{2} f(u_i^m) \;.
\end{equation}
Then it is evaluated and added to $u_i^{m}$ to obtain $u_i^{m+1}$:
\begin{equation}
u_i^{m+1} = u_i^m + \Delta t f(u_i^{m+1/2}) \;.
\end{equation}
This requires the predictor step to be evaluated at all spatial points before the full timestep can be completed.
The boundary condition is periodic; meaning, for example, for $n$ spatial points $i+1$ is actually $\text{mod} ((i+1), n)$.

\section*{Acknowledgments}

This material is based upon work supported by NASA under award No.~NNX15AU66A under the technical monitoring of Dr.~Eric Nielsen and Dr.~Mujeeb Malik.
We also gratefully acknowledge the support of NVIDIA Corporation with the donation of the Tesla K40 GPU used for this research.

\bibliography{GPUPaper}
\bibliographystyle{aiaa}

\end{document}